\documentclass[12pt]{article}
\usepackage{latexsym,amsmath,amssymb,amsbsy,graphicx}
\usepackage[cp1251]{inputenc}
\usepackage[T2A]{fontenc}
\usepackage[russian,english]{babel}
\usepackage[space]{cite} 

\begin{document}

{\bf Levels of Chromospheric and Coronal Activity in Sun-like Stars and Various Types of Dynamo Waves}

\bigskip

\centerline {$E.A. ~Bruevich^1, M.M. ~Katsova^1, D.D. ~Sokolov^2$}

\centerline {\it $^1 Lomonosov ~Moscow ~State ~University, Sternberg ~Astronomical ~Institute,$}
\centerline {\it Universitetsky pr., 13, Moscow 119992, Russia}\

\centerline {\it $^2 Lomonosov ~Moscow ~State ~University, Faculty ~of ~Physics,$}
\centerline {\it Vorob'evy gory, Moscow 119992, Russia}\

\centerline {\it e-mail:  {red-field@yandex.ru} }\

{\bf Abstract.} We analyze the X-ray emission and chromospheric activity of Sun-like stars of F, G, and K spectral classes (late-type stars) studied in
the framework of the HK project. More powerful coronas are possessed by stars displaying irregular variations
of their chromospheric emission, while stars with cyclic activity are characterized by comparatively modest
X-ray luminosities and ratios of the X-ray to bolometric luminosity $L_X/L_{bol}$. This indicates that the nature of
processes associated with magnetic-field amplification in the convective envelope changes appreciably in the
transition from small to large dynamo numbers, directly affecting the character of the ($\alpha$-- $\Omega$) dynamo. Due to the strong dependence of both the dynamo number and the Rossby number on the speed of axial rotation, earlier correlations found between various activity parameters and the Rossby number are consistent with our conclusions. Our analysis makes it possible to draw the first firm conclusions about the place of solar activity among
analogous processes developing in active late-type stars.

\bigskip

\vskip12pt
\section{Introduction}
\vskip12pt

 The first systematic investigations of solar-type
activity in other stars were begun in the middle of the
1960s by Wilson at Mt. Wilson Observatory, and have
continued since that time in the framework of the so called "HK project" [1, 2]. Observational studies of
chromospheric activity include determinations of the
ratio of the central fluxes of the Ca II H and K lines
(3968 \AA and 3934 \AA, respectively) to the flux in the
nearby continuum (4001 \AA and 3901 \AA)-- the quantity
$S_{Ca II}$(the mean ratio for the two lines). This approach
has provided a uniform description of levels of chromospheric activity and made it possible to monitor such activity over several decades.
Several dozen fairly bright stars with spectral types
from F2 to M2, in which it was believed solar-type
magnetic activity could develop, were chosen for study.
Beginning in 1977, the number of program stars grew
to 111, and this database became fundamental for
investigations of chromospheric variability on time
scales from decades down to the rotational periods of
the stars.
The quantity $S_{Ca II}$ (further S) proved to be a good
indicator of chromospheric activity [3]. Objects with
both large and small values of S are encountered among
stars with similar spectral types. Baliunas et al. [3]
attributed these differences in the level of chromospheric activity primarily to differences in age, at least
in the case of single stars. The mean (for the given interval of effective temperatures) value of S varies somewhat with spectral type.
It has now become clear that the processes giving
rise to the complex of activity phenomena manifest in
the interiors--convective zones--of stars encompass
virtually the entire atmosphere from the photosphere to
the corona. The first results on variations of the optical
continuum of stars in the HK project over roughly the
last ten years are presented in [4]; results for some
BY Dra stars (spotted red dwarfs) are presented in
[5, 6]. In general, these data testify that the photospheric activity of these stars resembles that of the Sun,
but, in some cases, the relative areas occupied by star
spots are two to three orders of magnitude larger than
on the Sun at the solar-cycle maximum. Data that can
be used to compare the properties of activity in different levels of the outer atmospheres of specific stars and
the Sun are already available.

\begin{figure}[tbh!]
\centerline{
\includegraphics[width=110mm]{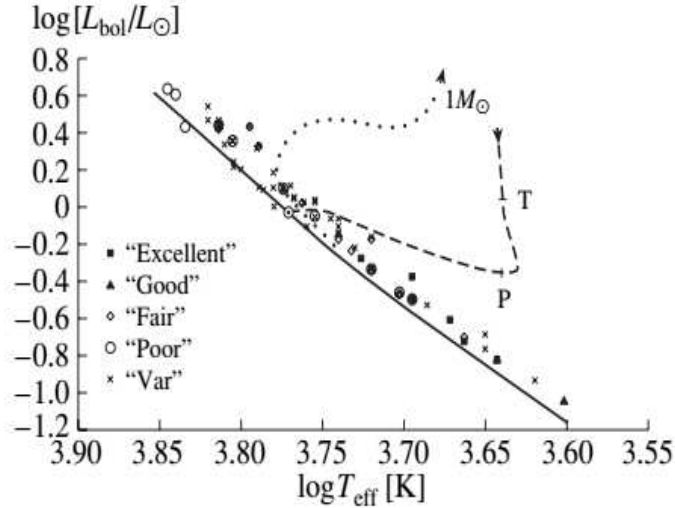}}
 \caption{Position in the Hertzsprung -- Russel diagram of late-type stars with chromospheric activity included in the HK
project. The various symbols correspond to different manifestations of chromospheric activity, from chaotic irregular 
variations (crosses) to cyclic activity (filled squares). The
solid curve shows the main sequence; the dashed curve
shows the track for the arrival to the main sequence of a
solar-mass star from the evolutionary status of T Tauri (T)
through the post-T Tauri stage (P). The dotted curve shows
the departure of a star of the same mass from the main
sequence. The position of the Sun is marked with a solar
sign. A number of stars with the same spectral type are
superposed, so that the number of stars in the diagram is
fewer than the number in Tables 1 - 4.}
{\label{Fi:Fig1}}
\end{figure}

\begin{figure}[tbh!]
\centerline{
\includegraphics[width=110mm]{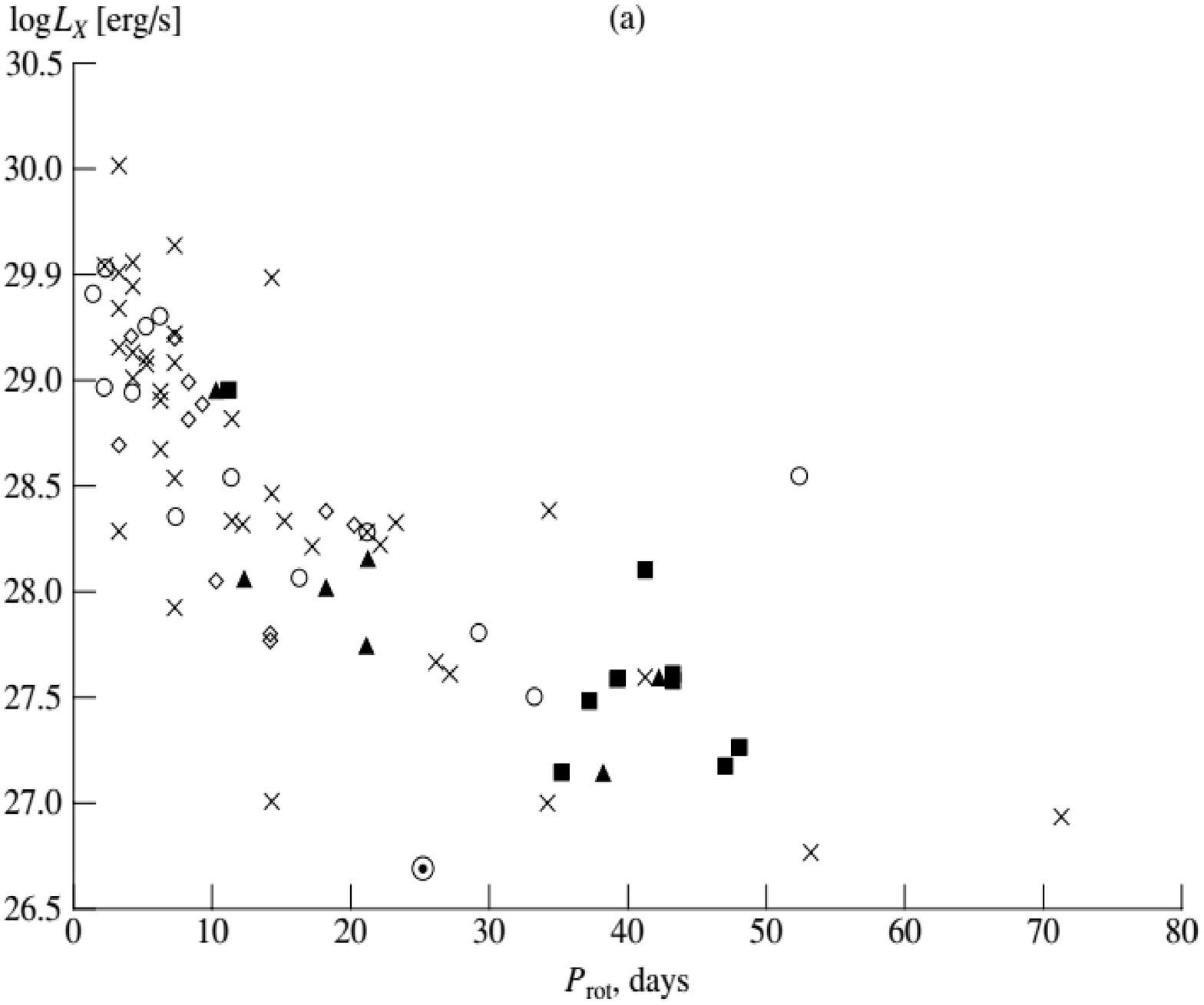}}
\centerline{
\includegraphics[width=110mm]{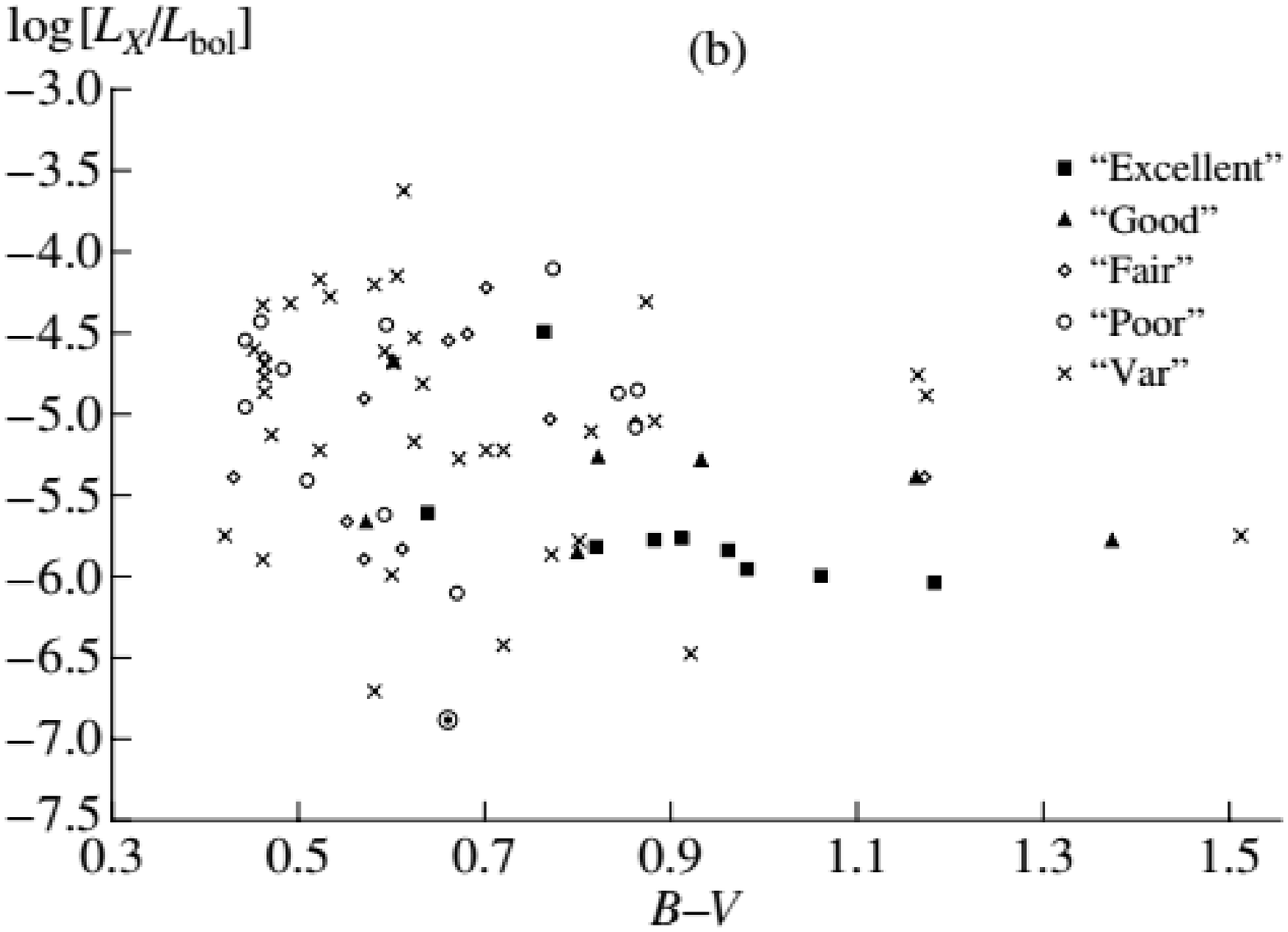}}
 \caption{(a) X-ray luminosity as a function of the period of axial rotation for HK project stars (from ROSAT data). (b) Ratio of the
X-ray to the bolometric luminosity for HK project stars as a function of B-V. Notation is the same as in Fig. 1.}
{\label{Fi:Fig2}}
\end{figure}

Long-term, uniform data sets on the coronal fluxes
of specific late-type stars are not yet available, except
for the Sun. However, measurements of the soft X-ray
emission of more than 1000 active late-type stars have
been made. These are primarily data obtained for nearby,
bright stars by the ROSAT satellite at 0.1-2.4 keV [7, 8].
The ROSAT data confirmed and expanded the general
results obtained by the Einstein observatory concerning
the dependence of the X-ray luminosity on the speed of
axial rotation, at least for main-sequence stars; the saturation of the X-ray luminosity relative to the bolometric luminosity $L_X/L_{bol}$ for red dwarfs (discovered by
Vilhu and Walter [9]); and the presence of high-temperature patches in the coronas of some late-type stars [10].
Important information has been obtained for active latetype stars in RS CVn binary systems, enabling in a number of cases a better understanding of the regularities
observed for active processes on single late-type stars.
An analysis of the X-ray emission of stellar coronas
based on Einstein data was presented by Katsova et al.
[11], who determined the densities at the bases of the
uniform coronas of dwarfs of various spectral types.
The coronas with the highest densities are possessed by
the most active red dwarfs, with spectral types K5-M3;
their coronal densities are an order of magnitude higher
than the corresponding solar value.
The X-ray emission of HK project stars detected by
ROSAT was considered by Hempelmann et al. [12],
who compared the X-ray emission of late-type stars
with constant levels of Ca II line emission and with regular and irregular (chaotic) long-term variations of this
emission. They found that the distribution of X-ray
fluxes associated with the stellar coronas is appreciably
different for stars with variable Ca II line fluxes: on
average, stars with irregular chromospheric activity
have higher X-ray fluxes. Hempelmann et al. [12]
attempted to relate this result to a dependence of coronal activity on the Rossby number, and to the possible
existence of a Maunder minimum at the current epoch
for some quiescent late-type stars.
Uniform X-ray catalogs of the nearest bright stars
are now available [7, 8], as well as the results of new
analyses of long-term chromospheric and photospheric
activity. 

Therefore, it is valuable to investigate the coronal activity of HK project stars with different types of
cycles and compare the results with the main expectations of dynamo theory. This can provide information
about the effectiveness of the dynamo mechanism on
F-K stars with differing levels of activity.

\vskip12pt
\section{Chromospheric activity
and position of a star
on the Hertzsprung -- Russell diagram}
\vskip12pt

\begin{figure}[tbh!]
\centerline{
\includegraphics[width=110mm]{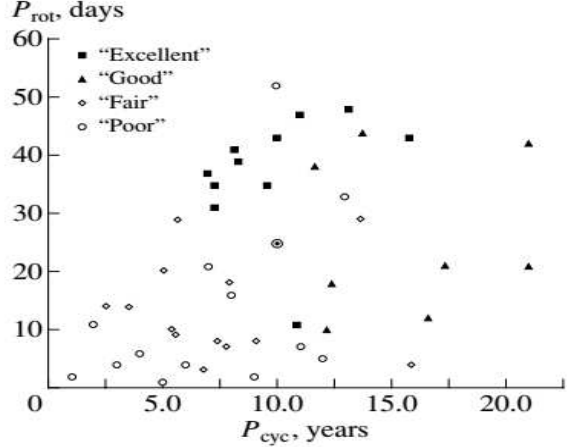}}
 \caption{Periods of axial rotation and cycle durations for HK project stars with more or less well defined activity cycles.
Notation is the same as in Fig.1.}
{\label{Fi:Fig3}}
\end{figure}

\begin{figure}[tbh!]
\centerline{
\includegraphics[width=110mm]{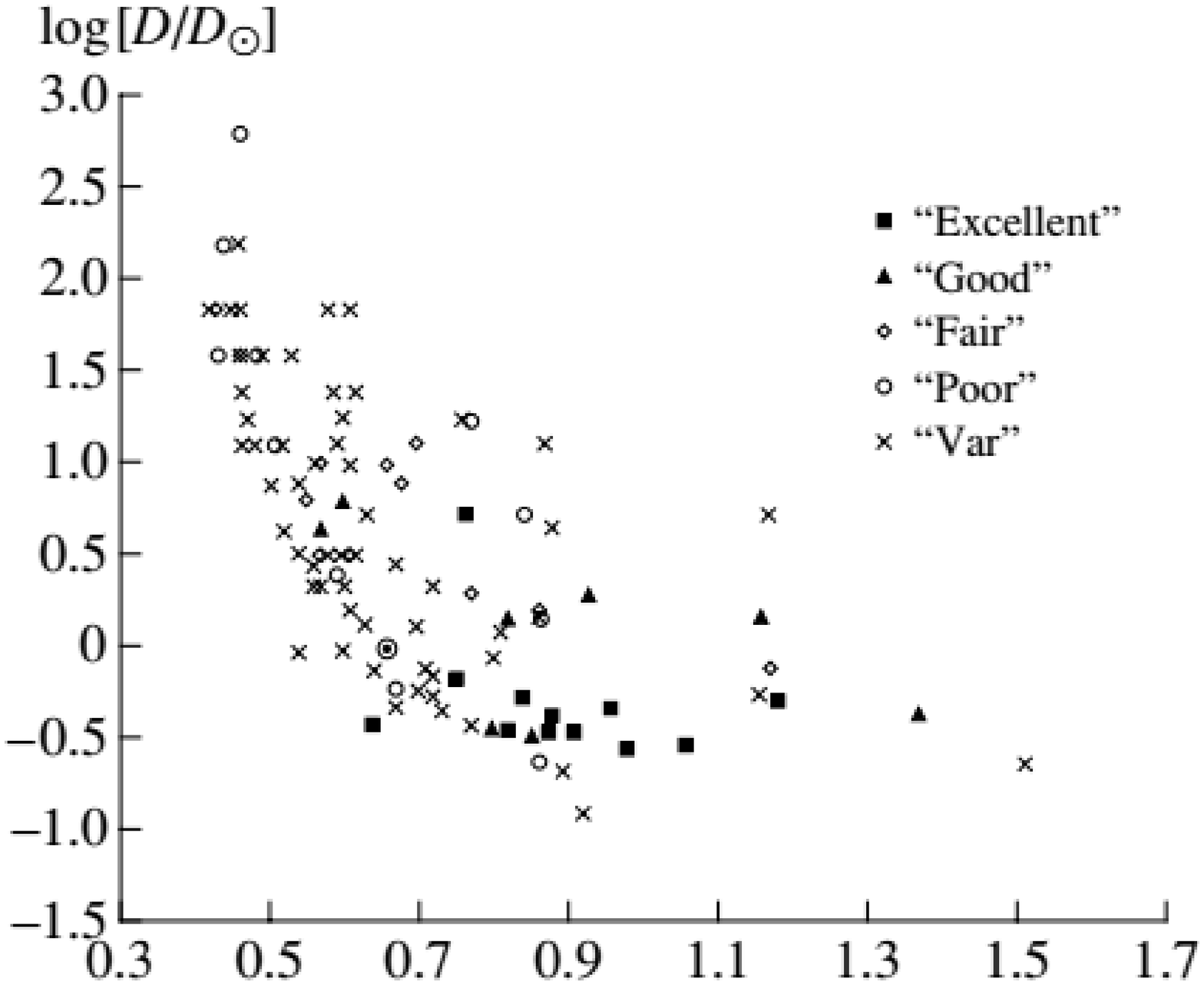}}
 \caption{Variation of the dynamo number (relative to the solar value) as a function of B-V for HK project stars. Notation
is the same as in Fig. 1.}
{\label{Fi:Fig4}}
\end{figure}

Chromospheric activity
and position of a star
on the Hertzsprung -- Russell diagram
In most stars with variable Ca II line emission, the
calcium-line flux shows modulations with the period of
rotation of the star. This rotational modulation virtually
disappears for stars of similar spectral types with the
smallest and largest values of S, due to either a lack of
active regions on these stars or, on the contrary, the
presence of a very large number of active regions more
or less uniformly distributed over the surface (if the
stellar rotational axis is perpendicular to the line of
sight, this implies a uniform distribution in longitude).
Observations carried out over the past 35 years have
enabled the detection of long-term variations of this
activity index for selected variable stars. Roughly one third of the studied stars show no periodicities in these
variations (this type of chromospheric variability is
designated after [3] as "Var"). Long-term periodicity in
variations of S were directly detected in 50 of the 111
HK-project stars. Applying spectral analyses to the
long-term data on chromospheric activity made it possible to determine for each star the degree to which cycles analogous to the solar cycle were present and
well defined. The periods for the cycles were from several years to decades.
Using formal statistical criteria, Baliunas et al. [3]
categorized stars into four groups according to the
degree to which their chromospheric-activity cycles
were well defined: "Excellent", "Good", "Fair" and
"Poor". This division was carried out based on the probability that the periodicity for stars in a given group was
not associated with Gaussian noise. A quantitative index
for this division is the false alarm probability (FAP), or
the probability that the observed behavior is the result
of Gaussian noise; for "Excellent" cycles FAP < $10^{-9}$,
for "Good" cycles $10^{-9}$ < FAP < $10^{-5}$, for "Fair" cycles
$10^{-5}$ < FAP < $10^{-2}$, and for "Poor" cycles $10^{-2}$ < FAP <
$10^{-1}$. A standard statistical treatment indicates that the
critical value of the FAP for which the null hypothesis
of the presence of cyclicity can be rejected is about
0.05. Formally, this means that one out of 20 cases that
are in fact due to Gaussian noise will be taken to represent cyclic behavior. However, between these extreme
cases of pure Gaussian noise and strictly periodic variations, there are many different possible types of variability that formally contain some periodic components, which can be revealed by Fourier analysis. In
some cases, it is also important to accurately exclude
harmonics. Therefore, a trustworthy identification of
cycles requires the application of not one but several
methods and criteria. Recent progress in finding solutions to this problem has come about with the development of wavelet analysis, which enables the analysis of
non-uniform time series. In summary, the FAP can
serve as a certain relative measure of the extent to
which cyclic variations are well defined. The broad
range for the "Good" group ($10^{-9}$ < FAP < $10^{-5}$) essentially means that there is already a watershed here
dividing cycles that can be distinguished with certainty
from cases with more complex activity on time scales
of several years.
In their analysis, Baliunas et al. [3] note that cycles
are more easily distinguished for K stars and for G stars
with small S values, while the chromospheric activity
of some F stars is higher, but more chaotic (cyclicity is
virtually absent). This result can be represented pictorially by plotting the HK project objects with chromospheric activity on a Hertzsprung -- Russell diagram. We
determined the stars’ bolometric luminosities using the
fundamental parameters given in [13, 14] (see Tables 1 -- 4). Figure 1 illustrates the behavior of the stars with
chromospheric activity by plotting their bolometric
luminosities (in solar units) $log L_{bol}/L_{\odot}$ ( as a function of
their effective temperatures $ T_{eff}$ . Here, we have
drawn the conventional zero-age main sequence and
two evolutionary tracks demonstrating the arrival to
and departure from the main sequence of solar-mass
stars [15].

We can see that these stars are located fairly close to
the main sequence. This reflects the fact that the youngest stars, with spectral types later than F0, and evolved
objects were not considered either in the HK project or
in the current paper. However, there is some scatter in
the ages of the stars studied, which may affect their
activity levels to some extent. Our independent analysis
and the analysis of [3] show that the cyclicity of longterm chromospheric variations begins to be manifest in
stars of spectral type G0, and is clearly defined in spotted K stars, while this activity is irregular in F stars.
Recent data on chromospheric activity and new observations of long-term variations in the optical continuum of
the HK project stars enable us to refine this conclusion.
Beginning with [16], the idea has been developed
that all these late-type stars can be divided into two
groups according to their activity levels. This division
becomes more obvious when considering the calcium
emission fluxes. In other words, in place of the index S,
the flux of a star in both calcium emission lines normalized to the bolometric luminosity, log $R^{'}_{HK}$ [17], was introduced as a measure of activity, and began to be used
in subsequent studies. In relations plotting log $R^{'}_{HK}$ as
functions of spectral type or other stellar parameters, two
groups can clearly be distinguished, with stars with well
defined cycles being less active. This tendency has been
traced in several studies, in particular in [4], which presents the results of ten years of optical-continuum observations of a number of HK project stars. It has also
become possible to compare long-term activity at the
chromospheric and photospheric levels, which increases
our ability to determine the extent to which variations
display well defined cyclicity.

Note that the difference in the activity levels of these
two groups of stars depends on their speeds of axial
rotation, which also indirectly reflects the influence on
rotation of several factors, first and foremost age. All
these considerations lead us to believe that cycles can
be distinguished with a high degree of certainty only for
stars in the "Excellent" group. We have already noted that
stars in the "Good" group have FAPs that cover a very
wide range (four orders of magnitude). This means that
this group includes both stars with well defined cycles
(with periods, however, that have not been strictly
maintained over the forty years of observations) and
stars with irregular variations of their chromospheric
and photospheric emission. Note that Baliunas et al. [3]
ascribe only 13 stars, including the Sun, to the "Excellent" group; well defined cycles are observed in only a
few of the 8 stars in the "Good" group.

\vskip12pt
\section{X-ray emission and cyclic activity
of late-type stars}
\vskip12pt

To analyze the coronal emission of late-type stars, we
used ROSAT observations of more than 1500 nearby,
bright main-sequence stars and subgiants, contained in
the two X-ray catalogs of Hunsch et al. [7, 8]. Both catalogs present measurements in the 0.1-2.4 keV range
carried out either as part of the ROSAT all-sky survey
or during directed observations. We were able to find
X-ray flux measurements for 78 of the HK project
stars. We used the data for these objects from [7, 8]
reduced to a unified system. In several cases when the
total X-ray flux from a binary system was registered
(HD 131156AB, HD 165341AB, HD 201091, and
HD 201092, HD 219834AB), we ascribed this flux
equally to both components if their speeds of axial rotation were essentially the same. In two such cases, we
divided the total flux of the binary in proportion to the
rotational speeds of the two stars. Note also that the
X-ray luminosity of HD 81809 may be overestimated,
since this is a spectroscopic binary.
The X-ray luminosities $L_X$ for the HK project stars
(when observed) are presented in Tables 1,2 (for stars with
more or less regular chromospheric activity) and Tables 3,4
(for stars with chaotic variability). These tables also list
a number of other physical characteristics of these
stars: their color indices, spectra, periods of axial rotation $P_{rot}$ [18], bolometric luminosities $L_{bol}$, etc. For the
Sun, we present the X-ray luminosity in the ROSAT band
for epochs of high activity ($L_X = 5 \cdot 10^{26}$ erg/s) [19].
Figure 2a presents the X-ray luminosities of the
HK project stars and the Sun together with characteristics of their chromospheric activity as functions of their rotational periods. The main tendency for more rapidly
rotating late-type stars to possess higher X-ray fluxes--
first discovered in Einstein data [20]--can clearly be traced here. Recall that the stars in Fig. 2 are near the
main sequence (Fig. 1), for which the relationship
between X-ray luminosity and rotational speed is
undoubted. The presence of some scatter among the
points, especially the deviation from the lower envelope
toward larger luminosities, is due to the fact that some
stars, even those near the Sun, are younger than the
Sun. The influence of age is manifest first and foremost
in the rotation speed: younger stars have shorter rotational periods. However, the shift of the points in Fig. 2a
due to the influence of age is not strictly along the
X axis; this becomes clear, for example, if we join the
points from [10] for G stars with different ages. This
means that, even if we consider a single rotational
period, younger objects are higher in this diagram than
stars that are older or the same age as the Sun.
The uniform data considered here clearly show a
result noted earlier in [12]: the X-ray luminosities of
stars with irregular, chaotic activity are higher than
those of objects with well established activity cycles.
One exception is the rapidly rotating star V2292 Oph
(HD 152391; $P_{rot}$ = 11 days), whose unusual chromospheric and photospheric variations were recently noted in [21].

Analysis of the X-ray data for the groups of stars
with different degrees of cyclicity shows that stars
within a single group tend to cluster in a log $L_X$ -- $P_{rot}$ 
diagram. This is especially true of the eight stars in the
Excellent group and two stars in the Good group with
clearly defined cycles. These stars, with spectral types
G7-K7 (except for HD 81809, whose spectral type is
close to that of the Sun), are clustered near log $
 L_x \approx 27.5$ and $P_{rot} \approx40$ days. They also occupy a very small
region in the log $ L_{bol} - P_{rot}$ diagram with log $ L_X/L_{bol} \approx$
-5.7.

Stars with other activity parameters, which rotate
more rapidly, and which have poorly defined cycles
demonstrate less strong clustering; their X-ray luminosities are factors of three to one hundred higher than
those of stars with better defined cyclicity.
Similar behavior can be seen in Fig. 2b, which plots
the ratio of the X-ray to the bolometric luminosity as a
function of the B-V color index. We can see that stars
with well defined cycles have roughly the same ratio log
$ L_X/L_{bol} \approx$ - 5.7. At the same time, for most stars with
irregular chromospheric activity, this ratio covers the
wide range log $ L_X/L_{bol} $ = -4.2 to -6.0. These values are
far from the limiting value for X-ray saturation for
active red dwarfs and components of RS CVn binary
systems, log $ L_X/L_{bol} \approx$ - 3.0.
The mean value of  log $ L_X/L_{bol}$ for the Sun is roughly
a factor of 30 lower than for most of the HK project stars considered here. This is primarily due to the different ages for these stars and the Sun. In addition, this
could reflect individual properties of the Sun's activity
compared to the analogous processes developing on
other stars. Note that the Sun is the most rapid rotator
among the stars with well defined cycles (with the
exception of V2292 Oph, discussed above).
It is interesting to ask what determines the duration
of the cycle of an individual star. The available observations made over 35 years contain only limited information about this question. Figure 3 compares the cycle
durations and rotational periods for the groups of stars
with differing degrees of cyclicity. In a number of
cases, a statistical analysis leads to poorly defined,
long-term variations being represented as a superposition (as a rule, of two) periodic components; clearly,
such cycle durations should be treated with caution.
Therefore, we have presented two pairs of values in Fig. 3
(the duration of the cycle and its harmonic) only for the
three stars HD 78366, HD 114710 ($\beta$ Com), and HD
149661 (V2133 Oph), whose main cycles are characterized as "Good".
Stars with well defined cycles rotate slowly, and the
durations of their cycles are from 7 to 17 years. In con trast to the earlier opinion that well defined cycles usually have durations of 10-11 years, Fig. 3 shows that
the scatter in the durations is appreciable. As noted
above, V2292 Oph has an exception cycle, as can also
be seen in this figure.
The durations of the cycles of rapidly rotating stars
with poorly defined periods for their long-term variations cover a wide range from 2.8 to roughly 20 years,
but nearly half of them have "cycles" shorter than
8 years. This provides evidence that decreasing the
rotational speed not only leads to the formation of a stable cycle, but that the duration of this cycle will exceed
the values typical of rapidly rotating stars.
The information discussed above can be used to
address the question of the place of solar activity
among similar processes occurring on late-type stars.
Here, we should bear in mind several important factors.
The Sun has the earliest spectral type in the group of
stars with well defined cycles. The chromospheric
activity of the Sun as a star is similar to the mean activity level for stars of this group, as is indicated by a comparison of the flux in the Ca II H and K lines ;
this can be seen in Fig. 8a of [4]. On the other hand, the
X-ray luminosity of the Sun is appreciably smaller than
those of other stars with cycles, as is clearly reflected by
the ratio of its X-ray to its bolometric luminosity. In
addition, the Sun rotates appreciably more rapidly than
the remaining stars with well defined cycles. It is possible that the Sun is located near the boundary where chaotic, irregular activity is transformed into more regular,
cyclic behavior. This could be an important argument in
searching for the origin of the Maunder minimum.
Thus, we can see that stars with well defined activity
cycles and those whose activity is fairly irregular form
two clusters in our diagrams. Well defined cycles correspond to longer durations of the activity cycles, longer
rotational periods, and lower X-ray fluxes; i.e., lower
magnetic-field strengths. This last correlation can also
be seen in the figure presented in [23], although it is not
specially discussed in that work.
This last conclusion may, at first glance, seem paradoxical from the point of view of dynamo theory, which
explains both the magnetic-field strength and the cycle
duration in the framework of a single mechanism for
the generation of the magnetic field. Below, we will demonstrate that this correlation can fit into the stellar dynamo mechanism.

\begin{table}
\caption{Stars with regular chromospheric actifity}
\begin{center}
\begin{tabular}{clclclclclclclcl}

\hline

HD& B-V& Spectrum&log$L_X$,erg/s&$P_{rot}$,d&$P_{cyc}$,y&log$L_{bol}$,erg/s&$D/D_{\odot}$\\ 

\hline
 & & &  "Excellent"& & & & & \\
Sun& 0.66& G2 V& 26.70& 25& 10.0& 33.58& 1.00 \\
4628&0.88& K4 (K2V)& 27.59& 39& 8.4& 33.33& 0.41\\
10476&0.84& Kl V& - & 35&9.6 &33.37& 0.51\\
16160& 0.98& K3 V& 27.26& 48& 13.2& 33.22& 0.27\\
26965&0.82& Kl V& 27.61& 43& 10.1& 33.38& 0.34\\
32147& 1.06& K5 V&27.18& 47& 11.1& 33.18& 0.28\\
81809&0.64& G2 V&28.10& 41& 8.2&33.59 &0.37\\
103095 &0.75& G8 VI& 31& 7.3& 33.44& 0.65\\
152391&0.76& G7 V &28.94 &11& 10.9&33.42& 5.17\\
160346 &0.96& K3 V& 27.48& 37& 7.0&33.25& 0.46\\
166620 &0.87& K2 V& 43& 15.8& 33.34&0.34\\
201091 &1.18& K5 V& 27.15& 35 &7.3& 33.08& 0.51\\
219834B &0.91& K2 V& 27.58& 43 &10.0& 33.31& 0.34\\
 & & &  "Good"& & & & & \\
3651 &0.85& K0 V& &44& 13.8& 33.36& 0.32\\
78366& 0.60& G0 V& 28.95& 10& 12.2& 33.60& 6.25\\
114710 &0.57& G0 V& 28.06& 12& 16.6& 33.72& 4.34\\
115404 &0.93& K1 V& 28.02& 18& 12.4& 33.28& 1.93\\
149661 &0.82& K0 V& 28.16& 21& 17.4& 33.38&1.42\\
156026 &1.16& K5 V& 27.75& 21& 21.0& 33.13& 1.42\\
201092 &1.37& K7 V& 27.15& 38& 11.7& 32.93& 0.43\\
219834A&0.80& G5 IV-V& 27.59& 42& 21.0& 33.40& 0.35\\
 & & &  "Fair"& & & & & \\
1835&0.66& G2 V& 28.99& 8& 9.1& 33.54& 9.77\\
18256 &0.43& F5 V& 28.69& 3& 6.8& 34.05& 69.44\\
20630 & 0.68& G5 V& 28.89& 9& 5.6& 33.50& 7.72\\
26913 &0.70& G8 V& 29.20& 7& 7.8& 33.49& 12.76\\
82885&0.77& G8 IV-V& 28.38& 18& 7.9& 33.42& 1.93\\
100180&0.57& F7 V& 27.77& 14& 3.6& 33.72& 3.19\\
154417&0.57& F8 V& 28.82& 8& 7.4& 33.72& 9.77\\
157856& 0.46& F5 V& 29.21& 4& 15.9& 33.93& 39.06\\
161239&0.65& G6 V& 29&5.7& 33.58& 0.74\\
165341A &0.86& K0 V& 28.31& 20& 5.1& 33.35& 1.56\\
187691& 0.55& F8 V& 28.05& 10& 5.4& 33.75& 6.25\\
190007 &1.17& K4 V& 27.81& 29& 13.7& 33.10& 0.74\\
190406 &0.61& G1 V& 27.80& 14& 2.6& 33.64& 3.19\\
\hline

\end{tabular}
\end{center}
\end{table}

\begin{table}
\caption{Table 1.(Contd.) Stars with regular chromospheric actifity}
\begin{center}
\begin{tabular}{clclclclclclclcl}

\hline
HD& B-V& Spectrum&log$L_X$,erg/s&$P_{rot}$,d&$P_{cyc}$,y&log$L_{bol}$,erg/s&$D/D_{\odot}$\\ 
\hline
 & & &  "Poor"& & & & & \\
3229 &  0.44  & F2 V (F IV)& 29.53 &2&4.9&33.0&156.25\\
37394 &0.84 &K1 V & 28.55 &11&3.6&33.37&5.17\\
76572 &0.43 &F3 V &&4&7.1&34.05&39.06\\
82443 &0.77 &K0 V &29.3&6& 2.8& 33.42& 17.36\\
111456 &0.46&F6 V &29.41&1&7.0&33.94&625.0\\
120136 &0.48&F7 V &28.95&4&11.6&33.87&39.06\\
155885 &0.86 &K1 V &28.28&21&5.7&33.35&1.42\\
176051 &0.59&G0 V &28.06&16&100&33.69&2.44\\
182101 &0.44&F6 V &28.97&2&5.1&34.0&156.25\\
188512 &0.86&G8 IV &28.56&52&4.1&33.35&0.23\\
194012 &0.51 & F5 V &28.36&7&16.7&33.82&12.76\\
206860 &0.59 &G0 V &29.25&5&6.2&33.69&25.00\\
224930&0.67&G3 V&27,51&33&10.2&33.52&0.57\\
\hline

\end{tabular}
\end{center}
\end{table}

\vskip12pt
\section{Parameters of stellar cycles
and the Parker dynamo}
\vskip12pt

The explanation of the origin of stellar cycles in the
framework of dynamo theory proposed by Parker in
1955 can be reduced to the following. The poloidal
component of the large-scale magnetic field of a star,
which can be expressed in terms of the azimuthal component of the vector potential of the magnetic field A, is
wound up by differential rotation, leading to the formation of an azimuthal component of the magnetic field B.
In turn, the helicity of the convective flows acts on the
azimuthal field component, giving rise to a poloidal
magnetic field and closing the cycle of self-excitation.
This process can be described quantitatively using the
Parker dynamo equations, which, in the simplest case,
have the form

$$   \frac{\partial A}{\partial t} = \alpha B + \frac{\partial^2 A}{\partial\theta^2 }~,                   ~~~~~~~~~~~~~~(1)  $$

$$   \frac{\partial B}{\partial t} = -DGcos\theta \frac{\partial A}{\partial \theta} + \frac{\partial^2 B}{\partial \theta^2}~,                   ~~~~~~~~~(2)  $$

where $\theta$ is latitude, and $$G = r^{-1}\frac{\partial \Omega}{\partial r}  $$ is the radial gradient
of the angular velocity in units of its maximum value.
The quantity $\alpha$ is measured in units of its maximum value. The dimensionless number D, called the
dynamo number, characterizes the intensity of sources
of magnetic field generation. This number is expressed
in terms of parameters characterizing the hydrodynamics of the convective envelope; in the simplest case, we
can use the approximate expression

$$   \frac{ D}{ D_{\odot}} = \Bigg( \frac{ R}{ R_{\odot}} \Bigg)^3 \Bigg( \frac{h_{\odot}}{h} \Bigg) \Bigg( \frac{V^{\odot}_{conv}}{ V_{conv}} \Bigg)^2 \Bigg( \frac{\Omega}{\Omega_{\odot}}\Bigg)^2  ~,                   ~~~~~~~~~~~~~~(3)  $$

where R is the radius of the zone of magnetic-field generation, h is the scale height, $\Omega$ is the angular velocity
of the stellar axial rotation, and $ V_{conv}$ is the convective
velocity. Here, when calculating G, we exchanged $\frac{\partial \Omega}{\partial r}$
with $\frac{ \Omega}{ r}$ and estimated $\alpha$ using the so-called Krause formula (see, for example, [24]).

When deriving the Parker dynamo equations, it is
assumed that the convective zone, indeed, forms a sort
of envelope. If the convective zone is very narrow, then
the term corresponding to radial diffusion of the magnetic field must be kept in equations (1) and (2). In the
case of very narrow, vanishingly thin convective zones,
this term becomes dominant and disrupts the dynamo
action, so that the magnetic field decays. If the convective zone occupies the entire star, the generation of
magnetic field is possible, but it has an appreciably different character than in the case of a convective envelope. In particular, the generation sources must usually
have appreciably higher intensities than in the case of a
convective zone with moderate thickness. The question
of the character and origin of the change in the type of
dynamo in the presence of a degenerate point radiative
core remains poorly studied theoretically. We emphasize that our stars have convective zones of moderate thickness.

Applying (3) to specific late-type stars, we should
note that the first three multiplicative factors on the
right-hand side compensate each other to a large extent.
For example, if we use classical models for the convective zones, the products of these three factors for stars
from F5 to M3 differ by only a factor of two. Thus, if
we do not consider the limiting cases of stars with very
thin convective zones and the latest M stars with complete convection, the dynamo number can be taken to
be proportional to the square of the angular velocity of
rotation of the star.
Traditionally, the behavior of periods of stellar
activity has been analyzed using the Rossby number.
Since both the dynamo numbers and the Rossby numbers of our stars are primarily determined by the angular velocity of rotation (relative to that of the Sun), the
two parametrizations are similar in practice. However,
conceptually, the dynamo number characterizes the
sources of field generation in the dynamo mechanism
and appears in the Parker equations explicitly, so that
we will use this quantity in our analysis.
We estimated the dynamo numbers using (3), and
present the dependences of parameters of the stellar
cycles on the dynamo number (in units of the solar
value) in Fig. 4. We can see that well defined cycles correspond to comparatively low dynamo numbers D and
modest magnetic-field strengths. When the dynamo
number increases, the magnetic-field strength also
increases, as seems natural (increased intensity of the
generator leads to a growth in the generated field). However, the degree of organization of the cycle decreases,
so that stars with very well defined cycles give way to
stars with higher, less constant levels of activity and
weakly defined cycles.

\begin{table}
\caption{Stars with irregular (chaotic) chromospheric actifity}
\begin{center}
\begin{tabular}{clclclclclclclcl}

\hline

HD& B-V& Spectrum&log$L_X$,erg/s&$P_{rot}$,d&log$L_{bol}$,erg/s&$D/D_{\odot}$\\ 

\hline

3443& 0.72& G5 V& &30&  33.77& 0.69 \\
3795& 0.70& G3 V& & 33&  33.49& 0.57 \\
6920& 0.60& F8 V& 29.47& 14&  33.66& 3.19 \\
9562& 0.64& G2 V&  &29&  33.59& 0.74 \\
10700& 0.72& G8 V& 27.00& 34&  33.47& 0.54 \\
10780& 0.81& K0 V& 28.34& 23&  33.40& 1.18 \\
12225& 0.62& G1 V&  28.46&14&  33.63& 3.19 \\
13421& 0.56& F8 V& & 17&  33.73& 2.16 \\
16673& 0.52& F8 V& 28.54& 7&  33.79& 12.76 \\
17925& 0.87& K0 V& 28.08& 7&  33.34& 12.76 \\
22049& 0.88& K2 V& 28.32& 12&  33.33& 4.34 \\
22072& 0.89& G7 V&  & 55&  33.32& 0.21 \\
23249& 0.92& K0 V& 26.95& 71&  33.30& 0.12 \\
25998& 0.46& F7 V& 29.54& 2&  33.93& 156.25 \\
26923& 0.59& G0 V& 29.21& 7&  33.69& 12.76 \\
29645& 0.57& G3 V& & 17&  33.72& 2.16 \\
30495& 0.63& G1 V& 28.83& 11&  33.61& 5.17 \\
33608& 0.46& F6 V& 29.15& 3&  33.93& 69.44 \\
35296& 0.53& F8 V& 29.44& 4&  33.78& 39.06 \\
39587& 0.59& G0 V& 29.08 & 5 &   33.69& 25.0   \\
43587& 0.61& G0 V&       &20&33.64& 1.56\\
45067& 0.56& F8 V&       &8 &33.73& 9.77\\
61421& 0.42&F5 IV-V&28.28&3 & 34.08&69.44\\
72905& 0.62& F7 V &29.11& 5 & 33.63& 25.00\\
75332& 0.49& F7 V &29.56& 4 & 33.87& 39.06\\
76151& 0.67&G3 V& 28.33& 15 & 33.52& 2.78\\
88335& 0.46&F6 V&       &5 & 33.93& 25.00\\
88737& 0.56&F5 V&       &8 & 33.73& 9.77\\
89744& 0.54&F6 V&       &9 & 33.77& 7.72\\
95735& 1.51&M2.1 Ve&26.78&53&32.50& 0.22\\
97344& 0.61&G0 V&       &8 &33.64& 9.77\\
100563&0.46&F5 V& 29.13& 4&33.47& 39.06\\
101501&0.72&G8 V& 28.21& 17&33.47&2.16\\
106516&0.46&F6 V& 27.93& 7&33.93& 12.76\\
107213&0.50&F8 V&      & 9&33.85& 7.72\\

\hline
\end{tabular}
\end{center}
\end{table}

\begin{table}
\caption{Table 3. (Contd.)Stars with irregular (chaotic) chromospheric actifity}
\begin{center}
\begin{tabular}{clclclclclclclcl}

\hline

HD& B-V& Spectrum&log$L_X$,erg/s&$P_{rot}$,d&log$L_{bol}$,erg/s&$D/D_{\odot}$\\ 

\hline

114378&0.45&F5 V+F5 V&29.34&3&33.95&69.44\\
115043&0.60&G1 V& 28.05&6&33.66&17.36\\
115383&0.58&F8 V& 29.51&3&33.70&69.44\\
115617&0.71&G6 V&       &29&33.48&0.74\\
124570&0.54&F6 V&       &26&33.77&0.92\\
124850&0.52&F7 IV&  29.63    &7&33.79&12.76\\
126053&0.63&G3 V&      &22&33.61&1.29\\
129333&0.61&G0 V& 30.01     &3&33.64&69.44\\
131156A&0.76&G8 V&28.90      &6&33.42&17.36\\
131156B&1.17&K4 V&28.34      &11&33.10&5.17\\
136202 &0.54&F8 IV-V&       &14&33.77&3.19\\
137107AB&0.58&G2 V+G2 V&28.41      &14&33.70&3.19\\
141004 &0.60&G0 V&27.66      &26&33.66&0.92\\
142373 &0.56&F9 V&      &15&33.73&2.78\\
143761 &0.60&G2 V&       &17&33.66&2.16\\
155885 &0.86&K0 V&28.28      &21&33.35&1.42\\
158614 &0.72&G8 IV-V V&       &34&33.47&0.54\\
159332 &0.48&F4 V&       &7&33.89&12.76\\
165341 &1.16&K6 V&28.38      &34&33.13&0.54\\
176095 &0.46&F5 V&29.00      &4&33.93&39.06\\
178428 &0.70&G4 V&28.23      &22&33.49&1.29\\
182572 &0.77&G8 IV&27.59      &41&33.42&0.37\\
185144 &0.80&K0 V&27.61      &27&33.40&0.86\\
187013 &0.47&F5 V&28.68     &6&33.90&17.36\\
190360 &0.73&G8 V&       &38&33.46&0.43\\
207978 &0.42&G6 IV&       &3&34.08&69.44\\
212754 &0.52&F0 V&        &12&33.79&4.34\\
216385 &0.48&F7 IV&       &7&33.89&12.76\\
217014 &0.67&G5 V&       &37&33.52&0.46\\

\hline
\end{tabular}
\end{center}
\end{table}

To explain this correlation, we turn to the properties
of Eqs. (1) and (2). They have solutions in the form of
travelling waves, called dynamo waves, whose propagation through the convective zone reflects the phenomenon of stellar activity. The dispersion relation linking the
complex Eigen number of these equations-i.e., the
growth rate and cycle frequency-and the dynamo
wavelength separates out the dynamo wave with the
maximum growth rate. It is natural to suppose that it is
this wave that determines the formation of waves of
stellar activity, and that precisely its parameters are
inherited at the stage of evolution of the magnetic field
when non-linear effects cease the exponential growth of
the dynamo-wave amplitude. It is not difficult to convince oneself that the wavelength and period of this distinguished dynamo wave decrease with growth in D.
Precisely this tendency can be seen in Fig. 4 (see [23]).

The decrease in the wavelength of the dynamo wave
with growth in the dynamo number for a fixed convective-zone radius means that more and more half-waves of magnetic activity begin to be present between the
poles and equator. It is natural to expect that, in the nonlinear regime, each of these half-waves lives a comparatively independent life; in the language of the theory
of dynamical systems, Eqs. (1) and (2) describing the
magnetic field as a distributed system with an infinite
number of degrees of freedom can be approximately
reduced to a certain dynamical system. The number of
variables (degrees of freedom) in this system is determined by the number of half-waves present in the direction of propagation of the dynamo wave. 

This effective
number of degrees of freedom increases with the
dynamo number. Similar phenomena are typical of a
wide circle of problems in non-linear dynamics. This
was first noted by Hopf in his formulation of a scenario
for the transition from laminar to turbulent flow. The
growth in the number of degrees of freedom occurs in a
discrete fashion, making it tempting to interpret the
existence of discrete groups with different characters of
stellar activity-though, of course, distinguished somewhat arbitrarily-as a reflection of real discreteness associated with an increase in the number of half waves.
An increase in the number of degrees of freedom of
a dynamical system usually leads to a qualitative
change in its behavior. In the case of a small number of
degrees of freedom (modest dynamo number), the nonlinear evolution of the dynamical system has the character of self-excited oscillations (the system has
extreme cyclicity). As the number of degrees of freedom grows, a regime of chaotic behavior in the form of
strange attractors usually arises. We suggest that the
numerous cases of aperiodic variations of stellar activity are associated with large dynamo numbers, with
their corresponding chaotic regimes. The transition
from extreme cyclicity to strange attractors and chaotic
behavior represents a modern scenario for the development of turbulent flow from laminar flow; at its base
lies a bifurcation that increases the dimension of the
corresponding dynamical system. We propose that the
appearance of discrete classes of stellar activity has
analogous origins. In other words, we propose to interpret the fact that stars with well-defined cycles display
comparatively low-intensity magnetic-field generation,
while stars with more intensive field generation do not
have well defined cycles, as observational confirmation
of a "cyclic behavior-chaotic behavior" bifurcation in
the stellar dynamo, associated with an increase in the
effective dimension of the system as the dynamo number is increased.
A possible alternative interpretation is that, in the
case of large dynamo numbers, the magnetic field in the
depths of the convective zone preserves cyclic behavior, but, for some reason, the link between the magnetic
field and surface activity tracers is disrupted. We
believe that this point of view encounters more difficulties, since the overall level of activity is increased when
the dynamo number is increased, and only its cyclic
character disappears.
In connection with our proposed interpretation, it is
valuable to consider the position of the Sun in these diagrams separately. Of course, the assignment of the Sun
to the ranks of stars with very well defined activity periods is due to the anomalously small distance of the Sun
from the Earth, and the associated exceptionally
detailed study of solar activity that has been possible. It
is well known that the intensities of different solar activity cycles can differ substantially; observing two
such cycles on a distant star, it is doubtful that we
would classify its cyclic activity as ideal. On the other
hand, to a first approximation, the Hale polarity law is
fulfilled on the Sun; i.e., data on sunspots specify in
each hemisphere only a single half-wave of solar activity. At the same time, data for other tracers of solar
activity specify additional waves, such as the polar
wave, visible in analyses of polar facculae. Finally, the
time dependence of solar activity reveals a well defined
periodic component with an added chaotic component
in the form of rare long-term drops in activity such as
the Maunder minimum. All these data are consistent
with the idea that the position of the Sun in our diagrams is near the edge of the region of well-defined periodic behavior.

\vskip12pt
\section{Conclusions}
\vskip12pt

Thus, among late-type stars, the soft X-ray luminosities and ratios $L_X/L_{bol}$ are substantially higher for stars
with irregular activity than for those with well defined
cycles. This is consistent with the result obtained earlier
in [12]. A detailed analysis for stars with reliably distinguished activity cycles indicates that, for comparatively
slowly rotating stars of spectral types later than G0, the
power of the corona decreases with the development of
well defined cyclicity. This conclusion does not pertain
to flare stars, for which a new effective mechanism for
coronal heating probably begins to operate, associated
with the occurrence of numerous weak flares (microflaring). The influence of age on the X-ray flux can be traced
in our sample, however it is not strong for these solarneighborhood stars, and cannot be responsible for the
observed variations of $L_X$ and $L_X/L_{bol}$.
Our conclusions are relevant for stars with spectral
types from F5 to M3, in which the thickness of the surface convective zones is neither very small, nor very
large compared to the stellar radius. This is consistent
with the idea that realization of the dynamo mechanism
requires the existence of an envelope in which the magnetic-field amplification can develop. It goes without
saying that efficient operation of the dynamo mechanism requires differential rotation and turbulent convection.
Our analysis demonstrates that the level of X-ray
emission is primarily determined by the parameter that
also determines the efficiency of the dynamo mechanism-the dynamo number. In the stars considered
here, the influence of factors connected with the structure
of their convective zones compensate each other to a large
extent, so that, in accordance with (3), the dynamo number
is determined by the square of the speed of axial rotation.
The transition from small to large dynamo numbers gives
rise to a change in the character of the dynamo process,
increasing its chaotic component.
We note especially that the general dependence of the
activity on Rossby number discovered earlier is also preserved, due to the strong influence of the speed of axial
rotation on both characteristic numbers-the Rossby and
dynamo numbers. The behavior we have found has a more
direct relationship to the theory of the ($\alpha - \Omega$) dynamo.
This behavior can be disrupted in stars with anomalously
weak differential rotation, such as members of close
binary systems, whose differential rotation is damped
by tidal forces.
In conclusion, we note that these results prompt us
to re-evaluate the position of solar activity among analogous processes occurring on other stars. If we do not
consider the exceptional star V2292 Oph, the Sun is located at the boundary of the region in Fig. 2a occupied by stars with cyclic activity, with the Sun rotating
more rapidly than the other stars with well defined
cycles. Note that the Sun's level of chromospheric
activity is fairly high: the ratio of the luminosity in
chromospheric emission to the bolometric luminosity is
roughly the same as that of other stars with well defined
cycles (see, for example, [4]). On the other hand, the
solar corona is appreciably weaker than the typical
coronas of active late-type stars: the solar ratio $L_X/L_{bol}$
is about $10^{-7}$, while this ratio for other active stars on
the lower part of the main sequence is several orders of
magnitude higher.

\end{document}